\documentclass[%
 reprint,
 amsmath,amssymb,
 aps,
]{revtex4-2}

\usepackage{graphicx}% Include figure files
\usepackage{dcolumn}% Align table columns on decimal point
\usepackage{bm}% bold math
\usepackage{ulem}
\begin{document}

\preprint{APS/123-QED}

\title{Susceptible-Infected-Susceptible Model with Mitigation on Scale-Free Networks}% Force line breaks with \\
%\thanks{A footnote to the article title}%
%Endemic States of an SIS Model with Mitigation on Scale-Free Networks
%SIS Epidemics with Mitigation on Scale-Free Networks: A Heterogeneous Mean-Field Study
\author{J. G. S. Delboni}
\email{joaogabrieldelboni@gmail.com}
 \affiliation{Escola de Artes, Ciências e Humanidades, Universidade de São Paulo,
Av. Arlindo Béttio 1000, 03828-000 São Paulo, Brazil}%Lines break automatically or can be forced with \\
\author{M. O. Hase}%
\email{mhase@usp.br}
\affiliation{Escola de Artes, Ciências e Humanidades, Universidade de São Paulo,
Av. Arlindo Béttio 1000, 03828-000 São Paulo, Brazil}
 %Authors' institution and/or address\\
 %This line break forced with \textbackslash\textbackslash
%}%

%\collaboration{MUSO Collaboration}
%\noaffiliation

%\author{Charlie Author}
% \homepage{http://www.Second.institution.edu/~Charlie.Author}
%\affiliation{
% Second institution and/or address\\
% This line break forced% with \\
%}%
%\affiliation{
% Third institution, the second for Charlie Author
%}%
%\author{Delta Author}
%\affiliation{%
% Authors' institution and/or address\\
% This line break forced with \textbackslash\textbackslash
%}%

%\collaboration{CLEO Collaboration}%\noaffiliation

\date{\today}% It is always \today, today,
             %  but any date may be explicitly specified

\begin{abstract}

We investigate infectious disease spreading on scale-free networks using a heterogeneous mean-field approach applied to the susceptible–infected–susceptible model, incorporating a mitigation factor. Individual heterogeneity is incorporated through a power-law distribution, while a mitigation factor accounts for behavioral responses and external effects that effectively reduce transmission from infected individuals. This mechanism, inspired by Malthus–Verhulst-type constraints, introduces a nonlinear saturation effect that encodes self-limiting dynamics in a tractable way. Analytical results are supported by stochastic simulations. We find that the mitigation factor induces a nontrivial behavior in the probability that a link points to an infected node, which develops a maximum at finite infection rates. In contrast, the overall prevalence remains a monotonically increasing function of the transmission rate. Additionally, the mitigation mechanism leads to an inversion in the dependence of epidemic observables on the degree exponent at sufficiently high transmission rates. While in the standard model smaller exponents yield higher endemic prevalence, in the modified model this trend reverses, with larger exponents producing higher prevalence and increased infection probability along network links.

%\begin{description}
%\item[Usage]
%Secondary publications and information retrieval purposes.
%\item[Structure]
%ou may use the \texttt{description} environment to structure your abstract;
%use the optional argument of the \verb+\item+ command to %give the category of each item. 
%\end{description}
\end{abstract}
%\keywords{Suggested keywords}%Use showkeys class option if keyword
                              %display desired
\maketitle

%\tableofcontents

\section{Introduction}

Historically, infectious diseases have been the leading cause of mortality worldwide \cite{murray1997,who2019}, and its mathematical modeling have drawn the attention of multiple fields within the scientific community, due to both its theoretical richness and its relevance to society. Classical approaches to epidemic modeling are typically based on compartmental descriptions, in which individuals are grouped into classes according to their disease status, and the dynamics are governed by the flux of the population between the compartments. In their standard formulation \cite{kermack1927}, these models rely on the homogeneous mixing hypothesis, which assumes that all individuals interact uniformly, allowing the contagion process to be described in terms of average interaction rates \cite{keeling2008}.

While this approximation yields important results, it neglects aspects of the complexity observed in real contact patterns. In many settings, interactions are not uniformly random nor equally likely across the population \cite{newman2018}. Such heterogeneity has been shown to play an important role in shaping epidemic dynamics, particularly in networks with broad degree distributions \cite{newman2018}.

A natural way to incorporate such heterogeneity is provided by the heterogeneous mean-field (HMF) approach, also known as the degree-based mean-field (MF) approximation. In this approach, nodes are grouped into compartments according to their degree, and nodes with the same number of contacts are treated as statistically equivalent \cite{barabasi2016}. Consequently, dynamical quantities such as the density of infected individuals are defined separately for each degree class. This approach can preserve analytical tractability for some models while capturing key features of complex network structure. 

In this work, we consider the susceptible–infected–susceptible (SIS) model. In scale-free (SF) networks, the heterogeneity of the degree distribution leads to the divergence of its second moment in the thermodynamic limit. Within the HMF model, this implies a vanishing epidemic threshold. 

To improve the realism of the model, we incorporate a common aspect observed in epidemics: a proportion of infected individuals tends to reduce their interactions, either voluntarily or through imposed isolation. This means that the contact network among the members of the system is altered, and this change can be incorporated through various approaches. Saturation mechanisms have been proposed to model reduced effective transmission in highly connected individuals, which can restore a finite
epidemic threshold even in SF networks \cite{lebowitz2004}. Alternatively, one may introduce an adjustable parameter that controls the likelihood of connections between individuals within the dynamics \cite{dias2021}. Finally, mitigation mechanisms can be incorporated based on the assumption that the probability of encountering an infected individual should decrease as the infected population increases \cite{kim2021}; this latter approach will be discussed below.

According to some models, population growth occurs exponentially without any restrictions, assuming an infinite abundance of resources or conditions to support such growth \cite{malthus1798}. To prevent this uncontrolled proliferation, the concept of carrying capacity \cite{verhulst1838} was introduced, establishing a maximum population limit for the system. This conceptualization has also been studied in the context of growing networks \cite{hase2016}. Although seemingly straightforward, the incorporation of this saturation mechanism profoundly alters the behavior of the system, transforming exponential growth into a logistic function. In the context of epidemic modeling, this saturation term effectively alters transmission dynamics by reducing the strength of interactions involving infected nodes.

The non-trivial result found after introducing the carrying-capacity mitigation factor in the Barabási–Albert (BA) model \cite{barabasi1999} is the appearance of a finite infection rate at which the probability $\Theta$ that a randomly chosen link points to an infected individual reaches its maximum \cite{kim2021}. This outcome is not observed in the original model \cite{pastor2001b}, where the function $\Theta(\lambda)$ increases monotonically with the transmission rate $\lambda$. Interestingly, despite this change, the overall prevalence remains a monotonically increasing function of the infection rate in both models. These results were established exactly for BA model and a natural extension is to examine SF networks, which encompass a wider class of models characterized by heterogeneity in degree distribution. 
Furthermore, since both empirical and computationally generated networks are finite, understanding how epidemic variables depend on system size is a fundamental issue. To address this, we perform stochastic simulations using the Gillespie algorithm \cite{gillespie76}, which exactly reproduces the continuous-time stochastic dynamics.

The outline of the paper is as follows. In Section II, we review the HMF model on uncorrelated SF networks. In Section III, we introduce the model with a mitigation factor, referred to as the modified heterogeneous mean-field model (mHMF), in which links involving infected nodes are systematically weakened. We extend all previously obtained analytical results to more general SF topologies. In Section IV, we study the stable state of the system via computer simulations, and in Section V we present our concluding remarks.

\section{Heterogeneous Mean-Field}

We consider the susceptible–infected–susceptible (SIS) epidemic model, where individuals are either susceptible or infected. Infection occurs through contact, while infected individuals recover spontaneously at a given rate, returning to the susceptible state without immunity. Before introducing the HMF formulation, it is useful to briefly recall the homogeneous mean-field model. Under the homogeneous mixing assumption, the time evolution of the density of infected individuals at time $t$, $\rho(t)$, can be written as \cite{boguna2003}

\begin{equation}
\frac{d\rho(t)}{dt} =  \lambda \langle k \rangle  [1 - \rho(t)]\rho(t) -\mu \rho(t) ,
\end{equation}
where $\mu$ is the recovery rate, $\lambda$ represents the spreading rate per contact and $\langle k \rangle$ is the average degree of the network. The epidemic threshold is given by $\lambda_c = \mu / \langle k \rangle$. This approach assumes that all individuals contribute equally to the spreading process. The steady-state non-trivial condition leads to $\rho \sim \lambda - \lambda_c$ when $\lambda > \lambda_c$ and to the absorbing state $\rho = 0$ for for $\lambda < \lambda_c$ \cite{keeling2008}.

Nonetheless, empirical evidence has highlighted that many contact networks are heavy-tailed \cite{liljeros2001}, reducing the explanatory power of the average degree alone. Consequently, fluctuations play an important role in defining epidemic properties \cite{barrat2008}. If a pathogen spreads on a network, nodes with a higher degree are more likely to encounter an infected individual and therefore become infected. Moreover, once infected, such nodes are also more likely to transmit the pathogen. This requires a mathematical model to account for the degree of each node as an intrinsic variable influencing disease transmission dynamics \cite{barabasi2016}.

The HMF model has been introduced independently at different times under various names \cite{kiss2017}. It first appeared as the proportional mixing model \cite{nold1980, hethcote1982, hethcote1984}, was later studied as social heterogeneity \cite{may1991}, and was subsequently applied to power-law networks under the name of dynamical mean-field reaction rate equation \cite{pastor2001a}. The same ordinary differential equation system was later derived in a different formalism, where partnerships are treated as dynamical variables, particularly in the modeling of sexually transmitted infections \cite{keeling2002}.

In the HMF model all nodes with the same degree are considered dynamically equivalent at any given time. Moreover, the network structure is assumed to evolve on a timescale much faster than the dynamical processes occurring on it, defining the annealed regime \cite{ashcroft1976, castellano2010}. In contrast to the quenched case \cite{inproceedings_wang2003, hethcote1984}, where the topology (i.e., the specific pattern of connections between nodes) is fixed, the annealed description relies on an averaged network representation that preserves the degree distribution. In this section, this approach is employed to analyze the SIS model on SF networks.

The system can be approached analytically by writing the equation governing the time evolution of the density of infected individuals over time. Given the different node degrees, we write the SIS model for each degree separately to account for connectivity fluctuations. This model represents a scenario where infected vertices independently attempt to infect each susceptible neighbor at the given rate \cite{ferreira2018}. By considering the relative density $\rho_k(t)$ of infected nodes with degree $k$, defined as the probability that a node with $k$ links is infected, we can write the dynamical mean-field equations as \cite{pastor2001a}

\begin{equation}
 \frac{\partial\rho_k(t)}{\partial t} = \lambda k [1 - \rho_k(t)] \Theta_k(t) - \mu \rho_k(t).
	\label{eq_hmf}
\end{equation}

\noindent The creation term considers the probability that a node with $k$ links is susceptible, $1-\rho_k(t)$, and becomes infected through one of its contacts. This probability is proportional to the spreading rate $\lambda$, the degree $k$, and the probability $\Theta_k(t)$ that a link emerging from a vertex of degree $k$ points to an infected node. This last quantity reduces, under the homogeneous mixing assumption, to the global density of infected nodes $\rho(t)$. In heterogeneous networks, however, $\Theta_k(t)$ is a nontrivial function that must account for the different degree classes and their connectivity patterns \cite{barrat2008}, since the fraction of infected neighbors can depend explicitly on the node’s degree $k$ and on time $t$ \cite{barabasi2016}. In the absence of degree correlations, $\Theta_k(t)$ is independent of $k$ and will therefore be denoted simply by $\Theta(t)$. Thus, the probability that a randomly chosen neighbor is infected can be written as \cite{pastor2001a}

\begin{equation}
     \Theta(t) =  \frac{1}{ \langle k \rangle} \sum_{k}{kP(k)\rho_k(t)},
	\label{eq_theta_hmf}
\end{equation}

\noindent where $P(k)$ is the degree distribution of the network and $\langle k \rangle$ represents the mean degree. Since $\rho_k(t)$ depends on $\Theta$(t), we obtain a self-consistency equation that allows us to determine $\Theta$. 

In the stationary regime $\Theta(t) = \Theta$ and $\rho(t) = \rho$ are independent of time. Setting $\partial\rho_k/\partial t = 0$ in Eq. (\ref{eq_hmf}) we obtain 

\begin{equation}
  \rho_k = \frac{k \lambda \Theta}{ 1 + k \lambda \Theta },
	\label{eq_steady}
\end{equation}

\noindent where we have considered $\mu = 1$, since we can always scale the time accordingly. This equation shows that the higher the node connectivity, the higher the probability of being in an infected state. The total fraction of infected individuals is the weighted sum of all infected $k$-degree nodes, that is

\begin{equation}
 \rho = \sum_k{P(k) \rho_k},
	\label{eq_pho}
\end{equation}

\noindent which expresses the endemic infection probability.

\subsection{HMF Model on Scale-Free Networks}

We consider a scale-free network with degree distribution $P(k) \sim k^{-\gamma}$. Denoting by $m$ the minimum degree, the explicit formula for the degree distribution is 

\begin{equation}
P(k) = (\gamma - 1)m^{\gamma - 1} k^{-\gamma} = (\alpha + 1)m^{\alpha + 1} k^{-\alpha - 2},
	\label{eq_pk}
\end{equation}

\noindent where we approximate the connectivity $k$ as a continuous variable and define

\begin{equation}
\alpha = \gamma - 2.
\label{alpha}
\end{equation}
This choice follows from the fact that, in the limit $N \to \infty$, the mean degree $\langle k \rangle$ converges only for $\alpha > 0$, which is the regime of interest in this work.

From (\ref{eq_theta_hmf}), (\ref{eq_steady}), and (\ref{eq_pk}) we can write $ \Theta $ as

\begin{equation}
  \Theta = \alpha \lambda \Theta m^\alpha \int_{m}^{\infty} \frac{dk}{k^{\alpha}(1 + \lambda k \Theta)}.
	\label{eq_theta2}
\end{equation}

\noindent If we set $ x = \lambda k \Theta $ and $ \epsilon = \lambda m \Theta$, we can write (\ref{eq_theta2}) as

\begin{equation}
  \Theta = \alpha \epsilon^\alpha I_1(\epsilon, \alpha),
	\label{eq4.85}
\end{equation}

\noindent where

\begin{equation}
  I_1(\epsilon, \alpha) = \int_{\epsilon}^{\infty} \frac{dx}{x^{\alpha}(1 + x)}.
	\label{eqn:equacao-exemplo17}
\end{equation}

We are especially interested in the behavior of the system close to the infection threshold. In this region, the integral above can be examined analytically.

We focus on SF regime, corresponding to $0 < \alpha \leq 1$. In the range $0<\alpha <1$, $ I_1(0, \alpha) $ remains finite. Therefore,

\begin{equation}
I_1(\epsilon,\alpha) =
I_1(0,\alpha) - \int_{0}^{\epsilon} \frac{dx}{x^{\alpha}(1 + x)}
\end{equation}

\begin{equation}
=
B(1-\alpha,\alpha)
-
\frac{\epsilon^{1-\alpha}}{1-\alpha}
+
\mathcal{O}(\epsilon^{2-\alpha}),
\end{equation}

\noindent where $B(a,b)$ denotes the beta function. For $\alpha>1$ the integral diverges as $\epsilon\to0$, and the leading behavior is obtained by isolating the singular contribution for $I_1(\epsilon,\alpha)$. A closely related analysis was presented in \cite{dias2021}. Here we only summarize the resulting expressions.

\begin{equation}
I_1(\epsilon, \alpha) =
\left\{
\begin{array}{ll}
    \frac{\pi}{\sin(\alpha \pi)} - \frac{\epsilon^{1 - \alpha}}{1 - \alpha} + \mathcal{O}(\epsilon^{ 2 - \alpha }) & \quad 0 < \alpha < 1 
    \\
    - \ln \epsilon + \epsilon + \mathcal{O}(\epsilon^2)   & \quad \alpha = 1 
    \\
    \frac{\epsilon^{1 - \alpha}}{\alpha - 1} + \frac{\pi}{\sin(\alpha \pi)} + \mathcal{O}(\epsilon^{2 - \alpha}) & \quad 1 < \alpha < 2
    \\
    \epsilon^{-1} + \ln \epsilon  + \mathcal{O}(\epsilon) & \quad \alpha = 2 
    \\
    \frac{\epsilon^{1 - \alpha}}{ \alpha - 1} - \frac{\epsilon^{2 - \alpha}}{\alpha - 2} + \mathcal{O}(1) & \quad \alpha > 2,
\end{array}
\right.
\label{eq4.847}
\end{equation}

\noindent where $\Gamma(1-\alpha)\Gamma(\alpha) = \frac{\pi}{\sin{( \alpha \pi)}}$ (for non-integer values of $\alpha$) was used.

Therefore, different behaviors emerge depending on the value of the degree exponent. Combining Eqs. (\ref{eq4.847}) and (\ref{eq_theta2}) we organize the asymptotic behavior for $\Theta$ according to $\alpha$.

i) $0 < \alpha < 1$:

\begin{equation}
   \Theta (\lambda) \simeq \left( \frac{ \alpha \pi }{ \sin(\alpha \pi) } \right) ^{\frac{1}{1 - \alpha}}  ( \lambda m )^\frac{\alpha}{1 - \alpha}.
	\label{eqn:equacao-exemplo20}
\end{equation}

ii) $\alpha = 1$ (BA model):

In this case, we can obtain a closed form for the integral, namely $I_1(\epsilon, 1) = \ln{\left( 1 + \epsilon^{-1} \right)}$. Considering again the non-trivial solution for $\Theta$, we obtain

 \begin{equation}
     \Theta (\lambda) = \frac{1}{\lambda m } \left( \frac{1}{e^{\frac{1}{\lambda m }} - 1 } \right)  \simeq \frac{1}{ \lambda m } e^\frac{-1}{\lambda m }.
   	\label{eqn:theta_barabassi}
\end{equation}

iii) $1 < \alpha < 2$:

For $\alpha>1$ the system exhibits a finite epidemic threshold, given by 

\begin{equation}
   \lambda_c = \frac{\alpha-1}{\alpha m}.
   \label{eq_limiar}
\end{equation}

\noindent In this regime

\begin{equation}
   \Theta (\lambda) \simeq \left( \frac{- \sin( \alpha \pi)}{\alpha (\alpha - 1)}
   \frac{ m}{(\lambda m)^\alpha}
   \right)^{\frac{1}{\alpha - 1}} \left( \lambda - \lambda_c \right)^\frac{1}{\alpha - 1}.
   \label{eqn:equacao-exemplo23}
\end{equation}

iv) $\alpha = 2$:

Here, we can also obtain a closed form for the integral, $I_1(\epsilon, 2) = \epsilon^{-1} -\ln{\left( 1 + \epsilon^{-1} \right)}$. Then, 

\begin{equation}
   \Theta (\lambda) \simeq 2\frac{\lambda_c}{\lambda^2 } \frac{(\lambda - \lambda_c)}{[-\ln(\lambda - \lambda_c)]}.
   \label{eqn:equacao-exemplo24}
\end{equation}

v) $\alpha > 2$

Finally, in this regime, we have that

\begin{equation}
   \Theta (\lambda) \simeq  \frac{\alpha - 2}{\alpha - 1} \frac{1}{\lambda^2 m} \left( \lambda - \lambda_c \right).
   \label{eqn:equacao-exemplo26}
\end{equation}

To obtain the overall density of infected individuals, we use Eqs. (\ref{eq_steady}) and (\ref{eq_pho}), leading to

\begin{equation}
   \rho =  (\alpha + 1) \epsilon^{\alpha + 1} \int_{\epsilon}^{\infty} \frac{1}{x^{\alpha+1}} \frac{dx}{(1+x)},
   \label{eqn:equacao-exemplo26}
\end{equation}

\noindent which, together with the previously determined values of $\Theta$, yields the endemic prevalence. The leading terms are presented in Table~\ref{tab:asymptotic}.

\section{Modified Heterogeneous Mean-Field}

In this section, we introduce a modified version of the HMF model by incorporating a mitigating factor. The motivation is to account for realistic influences that impact the contagion process by weakening contact with infected individuals, which tends to reduce their effective participation in the contact network as prevalence increases. 

This mitigation, inspired by the Malthus-Verhulst \cite{malthus1798,verhulst1838} model, introduces a non-linear saturation effect. Although simplified, this approach can represent several mechanisms observed in real-world epidemics, such as mobility restrictions, behavioral adaptation, or external effects. For instance, in institutional settings, policies can require symptomatic individuals to stay home, thus reducing the number of active contacts as infection levels increase, especially since schools or workplaces are more likely to implement or enforce such measures when prevalence is high. Similarly, while self-isolation is a behavior at the individual level and can occur independently of the prevalence, the overall proportion of individuals isolating can increase with higher case reports. Public awareness and collective behavioral adaptations can also intensify in response to rising case numbers \cite{ferguson2020}. The weakening scheme proposed can also represent reduced interaction as a result of the symptoms of the particular illness under consideration. These mechanisms described often act simultaneously and may reinforce one another. While such dynamics are complex and context-dependent, the inclusion of this saturation term offers a tractable way to encode self-limiting effects within transmission dynamics.

In this model, we replace (\ref{eq_theta_hmf}) by

\begin{equation}
   \Theta^\textit{mHMF} = \frac{1}{\langle k \rangle} \sum_{k} k P(k) \rho_k (1 - \rho_k),
   \label{eqn:thetamodificado}
\end{equation}

\noindent which is lower than the corresponding value in the original model. Since $0 \leq\rho_k \leq 1$ we have that $\rho_k (1 - \rho_k) \leq \frac{1}{4}$, so the probability $\Theta^\textit{mHMF}$ is bounded by $\Theta^\textit{mHMF}_\textit{max} = \frac{1}{4}$.

If we combine Eqs. (\ref{eq_steady}) and (\ref{eqn:thetamodificado}), we obtain

\begin{equation}
   \Theta^\textit{mHMF} = f(\Theta^\textit{mHMF}) =  \frac{1}{\langle k \rangle} \sum_{k} k^2 P(k) \frac{\lambda \Theta^\textit{mHMF}}{(1+\lambda k \Theta^\textit{mHMF})^2} 
   \label{eqn:f(theta)}.
\end{equation}

\noindent The epidemic threshold follows from the condition for the onset of a nontrivial solution, namely 

\begin{equation}
\left[\frac{\partial}{\partial \Theta} f(\Theta)\right]_{\Theta=0} = 1,
\end{equation}

\noindent which also results in $\lambda_c = {\langle k \rangle} /{\langle k^2 \rangle} = (\alpha - 1)/\alpha m$. Here $\langle k^2 \rangle$ stands for the second moment of the degree distribution. Therefore, in the modified model we have the same epidemic threshold as in the classical HMF formulation.

From (\ref{eqn:thetamodificado}) we can write

\begin{equation}
   \Theta^\textit{mHMF} = \alpha \epsilon^\alpha I_2(\epsilon, \alpha)
   \label{eqn:equacao-exemplo26},
\end{equation}

\noindent where

\begin{equation}
   I_2(\epsilon, \alpha) = \int_{\epsilon}^{\infty} \frac{dx}{x^\alpha (1 + x)^2}.
\end{equation}

\vspace{1em}

\noindent Here we can integrate by parts in order to obtain

\begin{equation}
   I_2(\epsilon, \alpha) = \frac{\epsilon^{-\alpha}}{1 + \epsilon} - \alpha I_1(\epsilon, \alpha + 1),
\end{equation}

\vspace{1em}

\noindent thereby enabling us to utilize the results of the previous section. The asymptotic behaviors of both models are summarized in Table~\ref{tab:asymptotic}. They share the same critical exponents, and for $\alpha \geq 2$ the result coincides with the homogeneous mixing approximation.

\begin{table*}
\caption{\label{tab:asymptotic}
\textbf{Asymptotic behavior of $\Theta(\lambda)$ and $\rho(\lambda)$ near criticality.}}
\begin{ruledtabular}
\begin{tabular}{ccccc}
$\alpha$ 
& $\Theta^{\mathrm{HMF}}$ 
& $\Theta^{\mathrm{mHMF}}$ 
& $\rho^{\mathrm{HMF}}$ 
& $\rho^{\mathrm{mHMF}}$ \\ 

\hline

$0<\alpha<1$
& $ \left( \frac{ \alpha \pi m^\alpha }{ \sin(\alpha  \pi) } \right) ^{\frac{1}{1 - \alpha}} \lambda ^{\frac{\alpha}{1 - \alpha}}$
& $\alpha^{\frac{1}{1-\alpha}}\Theta^{\mathrm{HMF}}$
& $ \frac{\alpha + 1}{\alpha}\left( \frac{ \alpha \pi m}{\sin(\alpha \pi)} \right)^{\frac{1}{1 - \alpha}} \lambda ^\frac{1}{1 - \alpha}$
& $\alpha^{\frac{1}{1-\alpha}}\rho^{\mathrm{HMF}}$ \\ [5pt]

$\alpha=1$
& $\frac{1}{\lambda m } e^{-\frac{1}{\lambda m}}$
& $e^{-1}\Theta^{\mathrm{HMF}}$
& $2e^\frac{-1}{\lambda m}.$
& $e^{-1}\rho^{\mathrm{HMF}}$ \\ [5pt]

$1<\alpha<2$
& $\left( \frac{- \sin( \alpha \pi)}{\alpha^2 \lambda_c }
   \frac{ 1}{(\lambda m)^\alpha}
   \right)^{\frac{1}{\alpha - 1}}(\lambda-\lambda_c)^{\frac{1}{\alpha-1}}$
& $\alpha^{-\frac{1}{\alpha-1}}\Theta^{\mathrm{HMF}}$
& $\frac{\alpha + 1}{\alpha} m \lambda \left( \frac{-\sin(\alpha \pi)}{ \alpha \pi \lambda_c } \frac{1}{(\lambda m)^{\alpha}} \right)^\frac{1}{\alpha - 1} \left( \lambda - \lambda_c \right)^\frac{1}{\alpha - 1} $
& $\alpha^{-\frac{1}{\alpha-1}}\rho^{\mathrm{HMF}}$ \\  [5pt]

$\alpha= 2$
& $2\frac{\lambda_c}{\lambda^2 } \frac{1}{[-\ln(\lambda - \lambda_c)]}(\lambda-\lambda_c)$
& $\dfrac{1}{2}\Theta^{\mathrm{HMF}}$
& $3\frac{\lambda_c}{\lambda^2 } \frac{1}{[-\ln(\lambda - \lambda_c)]}(\lambda - \lambda_c)$
& $\dfrac{1}{2}\rho^{\mathrm{HMF}}$ \\ [5pt]

$\alpha\ > 2$
& $\frac{\alpha - 2}{\alpha - 1} \frac{1}{\lambda^2 m}(\lambda-\lambda_c)$
& $\dfrac{1}{2}\Theta^{\mathrm{HMF}}$
& $\frac{\alpha + 1}{\alpha - 1}\frac{\alpha - 2}{\alpha} \frac{1}{\lambda} \left( \lambda - \lambda_c \right) $
& $\dfrac{1}{2}\rho^{\mathrm{HMF}}$ \\ [5pt]

\end{tabular}
\end{ruledtabular}
\end{table*}

In the HMF the probability $\Theta$ is a monotonically increasing function of the control parameter $\lambda$, and we have that $\Theta \to 1$ in the limit $\lambda \to \infty$. In contrast, the modified model exhibits qualitatively different behavior. Instead of increasing monotonically, $\Theta^\textit{mHMF}$ develops a maximum. For the specific case of the BA network ($\alpha = 1$) this peak probability can be obtained analytically, yielding \cite{kim2021}

\begin{equation}
\Theta^\textit{mHMF}_p = \frac{\sqrt{\lambda_p m}-1}{\lambda_p m},
\label{peak}
\end{equation}

\noindent where $\lambda_p$ is the spreading rate at the peak.
%This result is in good agreement with our stochastic simulation, shown in top curve of Fig. \ref{fig:theta}. (tirar ou colocar o ponto)

\section{Epidemic Simulation}

In this section, we implement an individual-level stochastic simulation using the Gillespie algorithm \cite{gillespie76} for both the usual and mitigated SIS model on SF networks.

When the population exhibits heterogeneity, nodes with a higher degree are more likely to encounter an infected individual and therefore become infected. Moreover,
once infected, such nodes are also more likely to transmit the pathogen. Therefore, for uncorrelated networks, the infection probability is proportional to $\Theta$, which reduces to the proportion of infected individuals in the homogeneous case. In this setting, each susceptible node \( i \) contributes with an infection rate 
\( \Lambda_i^{\textit{inf}} = \lambda k_i S_i \Theta \), while each infected node 
\( j \) contributes with a recovery rate \( \mu I_j \), where \( S_i \) and \( I_j \) 
are indicator variables, equal to 1 if node \( i \) is susceptible and node \( j \) 
is infected, respectively, and zero otherwise, and \( k_i \) represents the degree 
of node \( i \).

The simulations were performed on an uncorrelated annealed network within the HMF framework, which depends only on the degree distribution and does not require an explicit adjacency matrix. In this approach, it is not necessary to track the identity of specific links, since only the number of contacts of each individual determines its contribution to the infection process. In the simulations, $\Theta$ is computed dynamically from the system state. We adopt this simulation scheme because the modified model also incorporates a mean-field contribution in its formulation. The degree sequence was generated using inverse transform sampling, as an implementation choice to reproduce the prescribed degree distribution \cite{barabasi2016}. The dynamics is implemented using the Gillespie \cite{gillespie76} direct method, with infection and recovery events drawn according to their respective rates, which is $\Lambda^\textit{} = \Lambda^\textit{inf} + \Lambda^\textit{rec} = \sum_{i}^{N}{\lambda k_i S_i \Theta } + \sum_{j}^{N}{\mu I_j }$.

%The computer implementation simulates the epidemic process on an uncorrelated annealed network. Since the heterogeneous mean-field framework depends only on the degree distribution (which was obtained using inverse transform sampling), the network structure is not represented by an adjacency matrix. In this approach, it is not necessary to track the identity of specific links, as only the degree of each node determines its contribution to the infection process. Given the list of degree sequences $\{k_i\}$ drawn from the prescribed distribution, infection events are selected proportionally to node degree.

%The dynamics is simulated using the Gillespie direct method, where at each step infection and recovery events are drawn with probability proportional to their rates, and both the event type and the target node are determined simultaneously.

%\begin{figure*}
%\includegraphics{fig_2}% Here is how to import EPS art
%\caption{\label{fig:wide}Use the figure* environment to get a wide
%figure that spans the page in \texttt{twocolumn} formatting.}
%\end{figure*}

\subsection{Endemic Equilibrium}

The simulations were carried out on networks generated with a minimum degree $m = 3$ and the natural cutoff, given by $k_{\text{max}} = m N^{1/(\gamma-1)}$ \cite{barabasi2016}. In order to obtain the endemic equilibrium, all simulations were initialized with a fraction of randomly infected individuals, $\rho_0 = 0.2$, and evolved until the proportion of infected individuals reached a stationary regime. The choice of a relatively high initial density of infected nodes was made to minimize the probability of the system falling into the absorbing state, even in the supercritical regime, which arises as a finite-size effect due to the limited number of individuals and becomes increasingly relevant as the system size $N$ decreases. The equilibrium prevalence was then computed by averaging the steady-state values over a given time window and across independent network realizations. 

In our approach, we specify only the degree sequence, without explicitly constructing the edge set or imposing constraints on the pairing of stubs. In this sense, the description remains at the level of the configuration model ensemble, in which connections are assumed to be random \cite{bollobas1980}. Consequently, one does not expect correlations between node degrees, even when using the natural cutoff of the degree distribution. As discussed in \cite{catanzaro2005}, correlations typically emerge when additional constraints, such as the prohibition of self-connections and multiple edges, are enforced during network construction, thereby biasing the pairing of stubs. 

\begin{figure}[h]
\includegraphics[width=0.49\columnwidth]{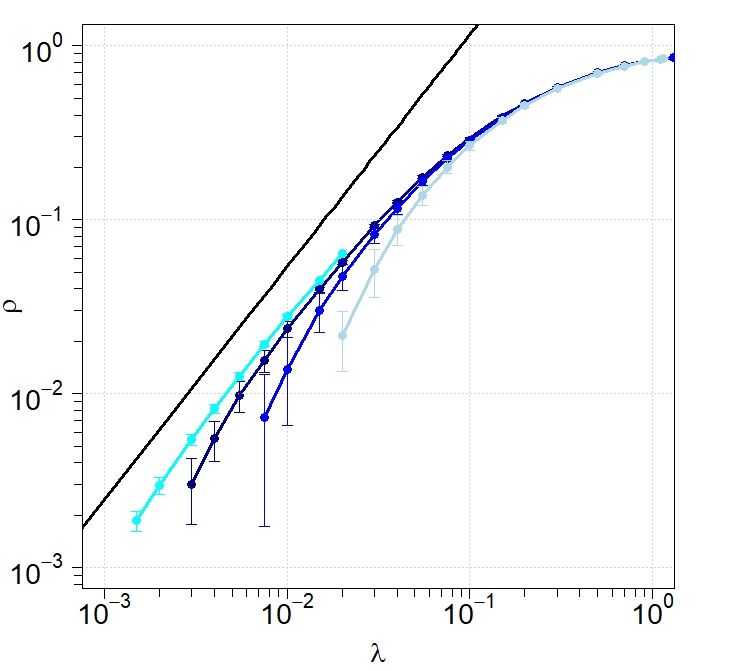}
\hfill
\includegraphics[width=0.49\columnwidth]{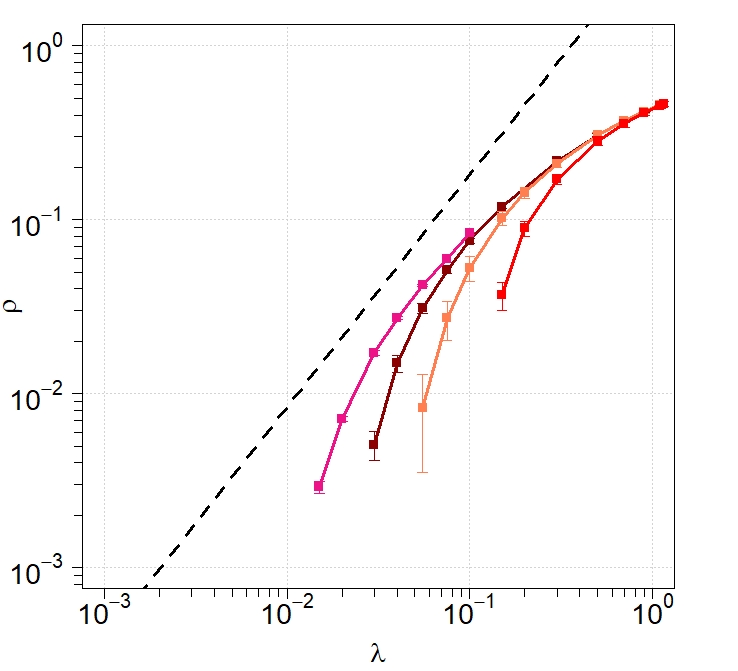}
\caption{\label{size_N} \textbf{Endemic prevalence as a function of the transmission rate $\lambda$ for different system sizes $N$.} From top to bottom: $N = 10^6, 10^5, 10^4,$ and $10^3$, with $\gamma = 2.25$. Circles (left) and squares (right) denote HMF and mHMF, respectively. Solid and dashed lines show the corresponding leading-order analytical results (Table \ref{tab:asymptotic}).}
\end{figure}

We performed simulations of both the HMF and mHMF models across different system sizes for $\gamma = 2.25$. In Fig.~\ref{size_N}, the behavior of the endemic prevalence $\rho$ as functions of the transmission rate $\lambda$ is shown. For sufficiently large values of $\lambda$, the curves corresponding to different system sizes collapse onto each other, whereas near the critical region, they become progressively shifted as $N$ decreases, reflecting finite-size effects. As the system size increases, the simulation results approach the leading-order analytical result. As expected, smaller systems are more prone to fluctuations, increasing the probability that low-density configurations reach the absorbing state. Finally, although finite systems display a threshold due to size effects, it can be observed that as the system size increases, the curves progressively approach the vanishing threshold predicted by the theory. The introduction of the mitigation factor appears to amplify finite-size effects, as deviations from the predicted critical exponent occur earlier compared to the original model.

\begin{figure}[h!]
\includegraphics[width=0.49\columnwidth]{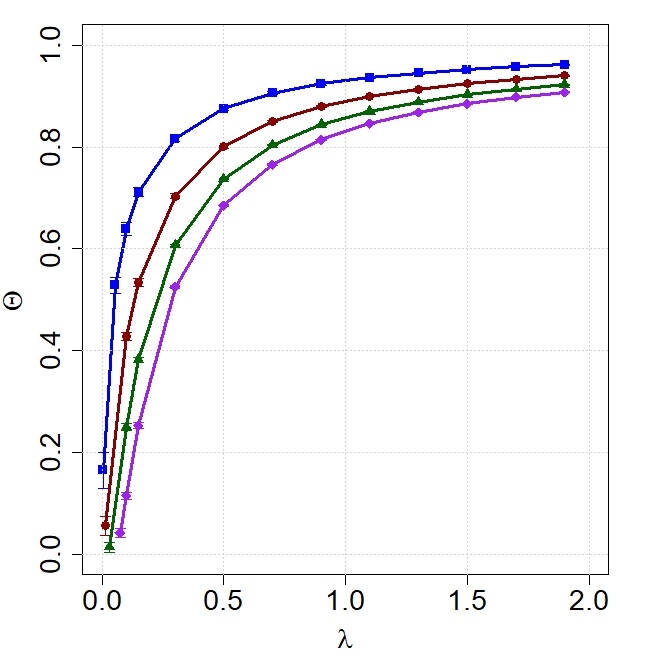}
\hfill
\includegraphics[width=0.49\columnwidth]{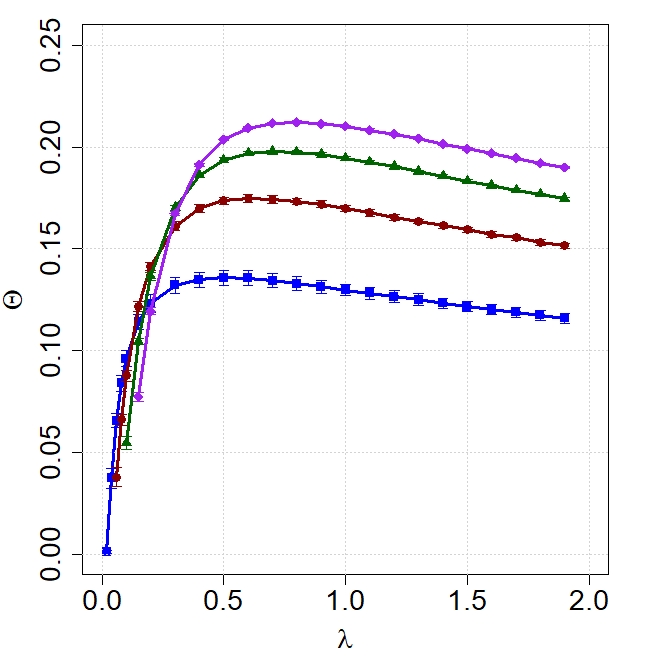}
\caption{\label{fig:theta} \textbf{Stationary $\Theta$ as a function of the transmission rate $\lambda$.} The network size is $N = 10^5$ and the degree exponents are 2.25 ($\square$), 2.50 ($\circ$), 2.75 ($\triangle$), and 3.00 ($\lozenge$). Left: HMF model. Right: mHMF model.}
\label{theta_fig}
\end{figure}

In Fig.~\ref{theta_fig} we show the endemic probability $\Theta$ for several exponents within the SF regime, with a fixed system size of $N = 10^5$ as a function of the transmission rate for both models.
In the standard model, $\Theta$ increases monotonically with the transmission rate. In contrast, the inclusion of the mitigation factor leads to a nontrivial behavior, with $\Theta$ no longer monotonic and instead displaying a maximum.

\begin{figure}[h!]
\includegraphics[width=0.49\columnwidth]{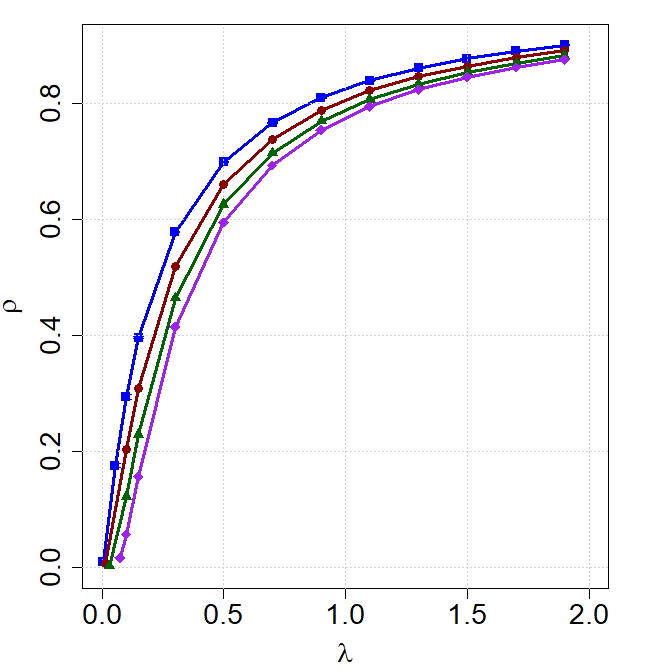}
\hfill
\includegraphics[width=0.49\columnwidth]{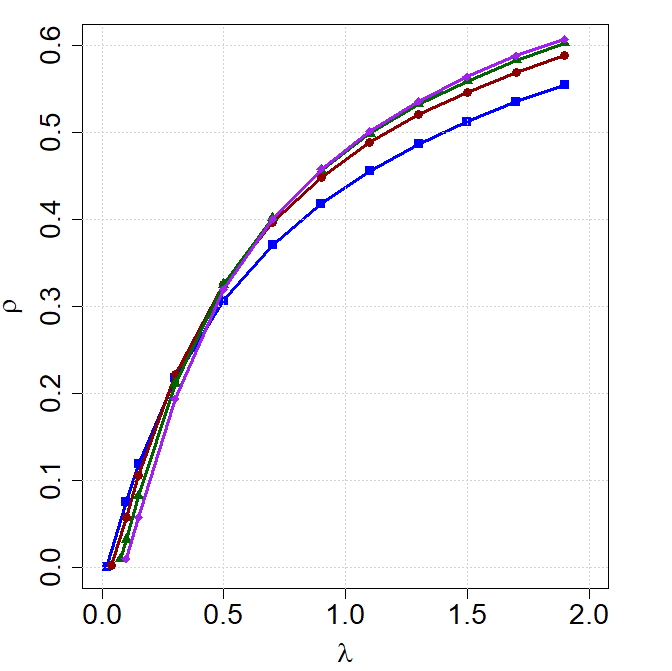}
\caption{\label{fig:theta} \textbf{Stationary $\rho$ as a function of the transmission rate $\lambda$.} The network size is $N = 10^5$ and the degree exponents are 2.25 ($\square$), 2.50 ($\circ$), 2.75 ($\triangle$), and 3.00 ($\lozenge$). Left: HMF model. Right: mHMF model.}
\label{rho}
\end{figure}

In Fig.~\ref{rho} we present the behavior of the prevalence $\rho$ as a function of the transmission rate for both models. In contrast with the non-monotonic behavior observed for $\Theta$, the prevalence remains a monotonically increasing function of the transmission rate, with quantitative differences between the models.

%For sufficiently high transmission rates, the curves exhibit a crossing behavior with respect to the degree exponent. 

Furthermore, the mitigated HMF shows a crossing between curves corresponding to different degree exponents. In the standard model, smaller exponents lead to higher prevalence, whereas in the modified model larger exponents yield higher values of $\Theta$ and $\rho$. This behavior can be understood intuitively by noting that, in the modified model, higher prevalence enhances the mitigation effect. As a result, scenarios that would otherwise sustain stronger spreading, such as those with higher transmission rates or larger effective connectivity due to the degree exponent, also experience stronger suppression, leading to the observed inversion.

%\begin{figure*}
%\includegraphics{fig_2}% Here is how to import EPS art
%\caption{\label{fig:wide}Use the figure* environment to get a wide
%figure that spans the page in \texttt{twocolumn} formatting.}
%\end{figure*}

\section{Concluding Remarks}

In this paper, we studied the SIS epidemic model on heterogeneous contact structures characterized by a scale-free degree distribution. We introduced a modified formulation that incorporates a mitigation mechanism imposing a saturation effect on the probability of transmission along links connected to infected nodes, inspired by Malthus–Verhulst model. This approach aims to capture behavioral responses and other external factors that reduce contacts involving infected individuals.

We derived exact analytical results and performed stochastic simulations using the Gillespie algorithm. The epidemic threshold and critical exponents remain unchanged with respect to the standard model. However, the mitigation mechanism leads to a qualitative difference. In particular, the probability that a link points to an infected node exhibits a maximum at finite transmission rates, while the global prevalence continues to increase monotonically. Furthermore, for sufficiently large transmission rates, we observe an inversion in the dependence of epidemic observables on the degree exponent. In the standard case, smaller degree exponents lead to higher prevalence. Here, we find the opposite behavior at high transmission rates, with larger exponents yielding higher prevalence and infection probabilities.

\pagebreak

%REV\TeX{} will automatically number such things as
%sections, footnotes, equations, figure captions, and table captions. 
%In order to reference them in text, use the
%\verb+\label{#1}+ and \verb+\ref{#1}+ commands. 
%To reference a particular page, use the %\verb+\pageref{#1}+ command.

%The \verb+\label{#1}+ should appear 
%within the section heading, 
%within the footnote text, 
%within the equation, or 
%within the table or figure caption. 
%The \verb+\ref{#1}+ command
%is used in text at the point where the reference is to %be displayed.  
%Some examples: Section~\ref{sec:level1} on %page~\pageref{sec:level1},
%Table~\ref{tab:table1},%

%\begin{acknowledgments}
%We wish to acknowledge the support of the author community in using
%REV\TeX{}, offering suggestions and encouragement, testing new versions,
%\dots.
%\end{acknowledgments}

%\appendix

%\section{Appendixes}

%\subsection{\label{app:subsec}A subsection in an appendix}

% The \nocite command causes all entries in a bibliography to be printed out
% whether or not they are actually referenced in the text. This is appropriate
% for the sample file to show the different styles of references, but authors
% most likely will not want to use it.
\nocite{*}

\bibliography{apssamp}% Produces the bibliography via BibTeX.

\end{document}